\documentclass[11pt]{article}

\begin{document}

\begin{center}
{\large \bf Gravitomagnetism, Frame-Dragging and Lunar Laser Ranging.}\\
\end{center}

\begin{centering}
Ignazio Ciufolini\\
\end{centering}

Dipartimento di Ingegneria dell'Innovazione, Universit\`{a} del Salento and INFN Sezione di Lecce, Via Monteroni, 73100 Lecce, Italy\\


Gravitomagnetism is a phenomenon predicted by Einstein's theory of
General Relativity, it consists in the interaction generated by a
mass current in a way formally similar to magnetism generated by an
electric current. A number of experiments have been proposed and
carried out to accurately measure the gravitomagnetic interaction.
The LAGEOS satellites observations \cite{ciup}, using the GRACE
models, have provided a measurement of the gravitomagnetic field
with accuracy of the order of 10 $\%$, the binary pulsar PSR 1913+16
has also provided an indirect astrophysical observation of this
field that might also be detected by refining the data analysis of
the Gravity Probe B space experiment \cite{muhl}. Frame-dragging is
a consequence of the gravitomagnetic interaction.

Recently, a number of papers have discussed whether the
gravitomagnetic interaction has already been accurately measured by
Lunar Laser Ranging (LLR) of the Moon orbit or not. This is a recent
chapter of a long debated question about the meaning of
frame-dragging and gravitomagnetism
\cite{ashs,ciu1,ciuw,ocon,mnt,kop,mnt2}. A basic question posed in
\cite{mnt,kop,mnt2} is whether the effect detected by Lunar Laser
Ranging is a coordinate, or frame-dependent, effect or not.

Here, we propose a distinction between gravitomagnetic effects
generated by the translational motion of a mass and those generated
by the rotation of a mass or by the motion of two masses (not
test-particles) with respect to each other. The geodetic precession,
or de Sitter effect, is a translational effect due to the motion of
the Earth-Moon gyroscope in the static field of the Sun, so it is
the effect discussed in \cite{mnt,mnt2}. The geodetic precession has
already been measured on the Moon orbit with accuracy of about 0.6
$\%$. The effect measured by LAGEOS and that might be measured by
Gravity probe B and LARES is due to the rotation of a mass, i.e., to
the rotation of Earth.

In order to distinguish between translational and rotational
gravitomagnetic effects we propose to use a spacetime invariant and
we show here that indeed what has been observed in \cite{mnt,mnt2}
is a translational gravitomagnetic effect. We stress that one cannot
derive rotational gravitomagnetic effects from translational ones,
unless making additional theoretical hypotheses such as the linear
superposition of the gravitomagnetic effects.

The translational gravitomagnetic effect, measured by LLR, depends
on the frame of reference being used in the analysis and is somehow
equivalent to the geodetic precession, the second effect, measured
by LAGEOS and LAGEOS 2, is an intrinsic gravitomagnetic effect
\cite{ciu1,ciuw} that cannot be eliminated by means of any
coordinate transformation.

In general relativity, in the frame in which a mass {\it M} is at
rest, we only have the nonzero components of the metric: $g_{00} = -
g^{-1}_{rr} = - \left(1 - {2 M \over r} \right)$, and $ g_{\theta
\theta} = g_{\phi \phi}$ sin$^{2} \theta = r^ {2} $, but we do not
have the so-called ``magnetic'' components $g_{0i}$. However, if we
consider an observer moving with velocity ${\bf v}$ relative to the
mass {\it M\/}, in his local frame we have ``magnetic'' components $
g_{0i} $ according to the formula $g'_{0 i} = \Lambda^ {\alpha}_{0'}
\, \Lambda^\beta_{i'} \, g_{\alpha \beta} \sim {M v^i \over r}$.
This ``magnetic'' component $ g_{0i} $ can simply be eliminated by a
Lorentz transformation back to the original frame, this effect has
been observed by Lunar Laser Ranging since the first measurements of
the geodetic precession of the Moon orbit. However, an object with
angular momentum $J$ generates a gravitomagnetic field intrinsic to
the spacetime structure that therefore cannot be eliminated by a
coordinate transformation, this is the field producing the
Lense-Thirring effect on the LAGEOS satellites. Indeed, in general
relativity, given explicitly a metric $g_{\alpha \beta}$, with or
without magnetic components $ g_{0i} $, in order to test for
intrinsic gravitomagnetism (i.e., not eliminable with a coordinate
transformation) one should use the Riemann curvature tensor $\bf R$
and the spacetime invariants built using it \cite{ciu1,ciuw}. In
\cite{ciuw} the explicit expression of the Riemann curvature
invariant $\bf {^\ast R \cdot R}$ is given, where $\bf {^\ast R}$ is
the dual of $\bf R$. For example, this invariant (really a
pseudo-invariant for coordinate reflections), no matter about any
coordinate transformation or any change of frame of reference, is
indeed different from zero in the case of the Kerr metric generated
by the angular momentum and the mass of a rotating body but is equal
to zero, as calculated in any frame and any coordinate system, in
the case of the Schwarzschild metric generated by the mass only of a
non-rotating body.

In \cite{ciuw}  it is shown that the gravitomagnetic effect on
LAGEOS and LAGEOS 2, due to the Earth angular momentum, is intrinsic
to the spacetime curvature and cannot be eliminated by a simple
change of frame of reference since the spacetime curvature invariant
$\bf {^\ast R \cdot R}$ is different from zero. However, below here
we show that the effect measured by Lunar Laser Ranging is just a
gravitomagnetic frame-dependent effect.

In \cite{mnt}  is shown that on the Moon orbit there is a
gravitomagnetic acceleration that  changes the Earth Moon distance
by about 5 meters with monthly and semi-monthly periods. This
variation of the Earth Moon distance is, in the Moon's equation of
motion, due to the term: $4 \sum_{j \neq i} \vec{v}_i \times
(\vec{v}_j \times \vec{g}_{ij})$, where, in the case of the Moon's
equation of motion, the indices i and j indicate Earth and Sun and
$\vec{g}_{ij} = {G M_j \over r^3_{ij}} \vec{r}_{ji}$ is the standard
Newtonian acceleration vector. The post-Newtonian point mass metric
generated by Sun and Earth has the non-diagonal term $g_{0i} = {7
\over 2} {M_\oplus v^i_\oplus \over r_\oplus}+ {1 \over 2} {M_\oplus
x^i_\oplus ({\bf x_\oplus \cdot v_\oplus}) \over r^3_\oplus} + {7
\over 2} {M_\odot v^i_\odot \over r_\odot}+ {1 \over 2} {M_\odot
x^i_\odot ({\bf x_\odot \cdot v_\odot}) \over r^3_\odot}$. In a
frame of reference comoving with the Sun we have the nondiagonal
gravitomagnetic term of the metric: $g_{0i} = {7 \over 2} {M_\oplus
v^i_\oplus \over r_\oplus}+ {1 \over 2} {M_\oplus x^i_\oplus ({\bf
x_\oplus \cdot v_\oplus}) \over r^3_\oplus}$, and by using the
geodesic equation of motion we find in this frame the Moon
gravitomagnetic acceleration $\sim \; \vec{v}_M \times
(\vec{v}_\oplus \times \vec{g}_{M \oplus})$, that is the term
discussed in \cite{mnt}. However, in a geocentric frame of reference
comoving with Earth, the nondiagonal gravitomagnetic term is $g_{0i}
= {7 \over 2} {M_\odot v^i_\odot \over r_\odot}+ {1 \over 2}
{M_\odot x^i_\odot ({\bf x_\odot \cdot v_\odot}) \over r^3_\odot}$
and the corresponding gravitomagnetic acceleration is: $\sim \;
\vec{v}_M \times (\vec{v}_\odot \times \vec{g}_{M \odot})$, that can
be rewritten as a part equivalent to the geodetic precession and
another one too small to be measured.

Therefore, the interpretation of the effect of Eq. (1) as a
gravitomagnetic effect in \cite{mnt} is clearly dependent on the
velocity of the frame of reference where the calculations have been
performed; for example one can start with the post-Newtonian
expression of the Schwarzschild metric generated by a mass
$M_\oplus$, then perform a Lorentz transformation with velocity
$v^i$ and have in the new frame the nondiagonal gravitomagnetic term
of the metric: $g_{0i} \sim {(M_\oplus v^i) \over r}$. From this
expression one can calculate the acceleration of the Moon due to the
mass $M_\oplus$ in this frame moving with velocity $v^i$ with
respect to the mass $M_\oplus$ and one then finds exactly the
acceleration $\sim \; \vec{v}_M \times (\vec{v} \times \vec{g}_{M
\oplus})$. This gravitomagnetic acceleration is then a
frame-dependent effect, indeed when we go back to the original
frame, where the mass $M_\oplus$ is at rest, this acceleration is
zero, unlike what happens in the case of the Lense-Thirring-Kerr
metric, where the Lense-Thirring effect by the Earth angular
momentum cannot be eliminated by a coordinate transformation.

This argument can be made rigorous by using the curvature invariant
${\bf ^\ast R \cdot R}$. This invariant is formally similar to the
electromagnetism invariant ${\bf  ^\ast F \cdot F}$ equal to ${\bf
E} \cdot {\bf B}$. In the case of a point-mass metric generated by
Earth and Sun, this invariant is: $\sim \bf G \cdot \bf H$, where
$G$ is the standard Newtonian electric-like field of Sun and Earth
and $H$ the magnetic-like field of Sun and Earth; this magnetic-like
field is $\sim {\bf v} \times {\bf G}$ and then clearly, on the
ecliptic plane, the invariant ${\bf ^\ast R \cdot R}$ is null.
Indeed, its precise expression, as calculated in a frame comoving
with the Sun, is: ${\bf ^\ast R \cdot R} \sim {M_\oplus M_\odot
\over {r^4_M \, r^4_{M \oplus}}} z_M (v^y_\oplus x_\oplus -
v^x_\oplus y_\oplus) (\hat{x}_M \cdot \hat{x}_{M \oplus})$, where
$z_M$ is the distance of the Moon orbit from the ecliptic plane;
this expression is then $\cong 0$ on the ecliptic plane (even by
considering that the Moon orbit is slightly inclined of 5 degrees on
the ecliptic plane, its z component would only give a contribution
to the change of the radial distance of less than 1 $\%$ of the
total change).

\end{document}